# Forecasting strong aftershocks in earthquake clusters from northeastern Italy and western Slovenia

S. Gentili[a] and R. Di Giovambattista[b]

[a] National Institute of Oceanography and Applied Geophysics - OGS, Italy
[b] Istituto Nazionale di Geofisica e Vulcanologia, Rome, Italy

## Abstract

In this study, we propose an analysis of the earthquake clusters that occurred in North-Eastern Italy and western Slovenia from 1977 to today. Given a mainshock generating alarm in the population, we are interested in forecasting if a similar magnitude earthquake will follow. We classify the earthquake clusters associated with mainshocks of magnitude $M_m$ into two classes: if the strongest aftershock has a magnitude $\geq M_m$-1 (swarms or large aftershock seismic sequences) as type A, otherwise (smaller aftershocks seismic sequences) as type B. A large aftershock following a main shock can cause significant damages to already weakened buildings and infrastructures, so a timely advisory information to the civil protection is of great interest for effective decision-making. For the first time, we applied to a new catalogue a pattern recognition algorithm for cluster type forecasting that we developed for all Italy (Gentili and Di Giovambattista, 2017). Thanks to the lower completeness magnitude of the local OGS catalogue, compared to the national one, and to a new version of the algorithm, we were able to lower the threshold of the clusters mainshocks magnitude from 4.5 to 3.7. The method has been validated by rigorous statistical tests. We tested the algorithm on the 1976 highly destructive earthquake cluster (mainshock magnitude 6.5 – the strongest in the last 80 years in the region) and we retrospectively forecasted it as an A cluster. Successful results were obtained also on other three smaller earthquake clusters in 2019.

1. Introduction

The Northeastern Italy and Western Slovenia correspond to the Southeastern sector of Alpine Chain and Northern Dinarides. The analyzed region is included in a convergent margin zone between the Adria microplate and the Eurasian plate (see e.g. Castellarin et al., 2006; Burrato et al., 2008). Recent seismological and space geodetic studies evidenced the existence of an Adriatic microplate moving independently relative to both Eurasia and Africa.

A Large part of the Adriatic microplate is deformed at different rates along three subduction zones, with orogens formation: it overrides the European plate along the Alps and subducts beneath the Dinarides and Apennines, where a complicated mechanism of roll-back is active (Cuffaro et al., 2010).

The undeformed part of the Adriatic plate is very limited and mainly covered by the Adriatic Sea, however it has been possible to estimate an anticlockwise rotation of Adria with respect to Eurasia by GPS observations (D'Agostino et al. 2005; Devoti et al. 2008).

The dominant mode of deformation changes from compressional to shear when moving from west to east. Several tectonic phases in the region inherited and reactivated the main pre-existing faults and fragmented the crust into different tectonic domains corresponding to different seismotectonic zones (Bressan et al., 2003). Seismicity concentrates along the almost E-W trending Alpine fault system and along the NW-SE trending Dinaric fault system. Earthquakes are shallow, in the range 6-12 km in the western part (Bressan et al. 2012, Bressan et al., 2018) and in the range 2-7 km in the eastern part (Bressan et al. 2009, Gentili and Gentile, 2015, Gosar, 2019). Fault plane solutions are prevalently of thrust type in the western part and strike-slip in the eastern one, with minor normal faulting events (Bressan et al., 2016 and references therein).

The region is one of the most seismically active in Italy and surroundings and was characterized by strong earthquakes in the past (1117 January 3, Veronese IMCS=IX-X; 1323, Novi Vinodolsk IMCS=IX, 1348, January 25, Carnia IMCS=IX-X; 1511 March 26, Slovenia IMCS=X;1695 February 25, Asolano IMCS=IX-X; 1721 January 12, Rijeka IMCS=IX; 1873 June 29, Bellunese IMCS = IX-X; 1928, March 23, Carnia, IMCS=IX, 1936 October 18, Bosco del Cansiglio IMCS=IX; 1976 May 6, Friuli IMCS=IX-X – see e.g. Gentili and Franceschina, 2011 and Fig. 1) causing death and destruction. Soon after the M=6.4 1976 earthquake (Slejko et al., 1999, Rovida et al., 2016), which caused about 1000 casualties and the complete destruction of little towns and villages, a seismic network managed by OGS (National Institute of Oceanography and Experimental Geophysics) was installed. From 1977, seismicity has been recorded by the OGS network. Main events, with magnitude greater than 5, recorded by the network are: the ML=5.2 16 September 1977 Trasaghis earthquake, in the central-western part of the analysed area, corresponding to Friuli Venezia Giulia Italian region, and two more recent earthquakes respectively on 12 April 1998 (ML=5.6) and on 12 July 2004 (ML=5.1), located near the border between Italy and Slovenia. However, several clusters of lower magnitude have been detected in different parts of the analyzed region.

This study is possible thanks to the OGS bulletins, available at the website www.crs.ogs.trieste.it/bollettino/RSFVG/RSFVG.en.html, an accurate local catalogue characterized by low completeness magnitude that has been compiled by the National Institute of Oceanography and Experimental Geophysics, Centre of Seismological Research, since 1977. Even if many clusters in the region are characterized by a mainshock followed by smaller aftershocks, in some cases repeated strong earthquakes have been recorded. These earthquakes constitute a big danger from the civil protection point of view, because they imply accumulated damage on already damaged structures, increasing the probability of building collapse. The last strong earthquake in the region, the 1976 Friuli earthquake, was followed four months later by similar magnitude earthquakes, causing a large toll in terms of casualties and destruction.

This paper is divided into two parts: a first part is a statistical analysis of the strongest aftershock characteristics (time, magnitude etc.). The second part applies a pattern recognition approach to estimate the probability of a strong following earthquake. Other methods have been previously proposed in literature for strong following earthquake forecasting. After the first pattern recognition study of Vorobieva and Panza (1993) and Vorobieva (1999), other works followed, like e.g. the Gutenberg Richter b-value based paper of Shcherbakov and Turcotte (2004), the energy and stress drop based paper by Gentili and Bressan (2008 - further details on relation between stress drop and released energy in Bressan et al., 2017) the pattern recognition method proposed by Gentili and Di Giovambattista (2017) and, in 2019, the b-value based paper of Gulia and Wiemer (2019) and the ETAS based study of Shcherbakov et al. (2019). The method proposed in this paper (here and after NESTORE – Next STrOng Related Earthquake) is a pattern recognition technique developed in its previous version by Gentili and Di Giovambattista (2017), using statistical features based on the number of the early aftershocks and on the spatio-temporal evolution of the radiated energy in the first hours/days after the mainshock. We performed the analysis on different timespans after the mainshock to simulate the increase of available information as time passes during the earthquake clusters. If the difference in magnitude between the mainshock and the strongest aftershock is lower than 1, clusters are classified by NESTORE algorithm as "Type A", otherwise as "Type B".

## 2. Earthquake catalogue and cluster detection

### *2.1. Catalogue and selected area*

The adopted catalogue is the OGS (Istituto Nazionale di Oceanografia e Geofisica) catalogue, recorded by the OGS network operating from 6$^{th}$ May, 1977 to today and interrupted only from 4$^{th}$ December 1990 to 21$^{st}$ May 1991 due to a fire accident (Renner, 1995; Priolo et al., 2005; Gentili et al., 2011, Bragato et al., 2013).

During more than 40 years of activity the OGS network sensitivity and the monitored area changed, depending on the evolution of the network geometry, the instrumental characteristics, the monitoring and processing strategies over time, and the collaborations with surrounding networks. In Gentili et al. 2011, the sensitivity has been assessed in terms of completeness magnitude by the Probability-based Magnitude of Completeness (PMC) method by Schorlemmer and Woessner (2008). The paper shows that it is smaller or equal to 1 in the central Friuli region, while it decreases and is more variable in surrounding regions. From 2000, the sensitivity increased westward in Veneto due to the higher number of seismic stations installed (Gentili et al., 2011) and from 2008 eastward in Slovenia, thanks to the collaboration with Slovenian and Austrian seismic networks (Peruzza et al., 2015).

In order to define the region of sufficient completeness of the catalogue, we started from the results of Peresan and Gentili (2018) for the analysis of the catalogue by the nearest neighbor method. They used the general scheme proposed by Kossobokov et al. 1999, in which they compared the ISC catalogue with the OGS one on 0.25x0.25 degrees cells in the time interval 1977-2015. The area of sufficient completeness is detected based on the ratio R between the number of ISC earthquakes that have an equivalent in OGS catalogue and the total number of earthquakes in ISC catalogue.

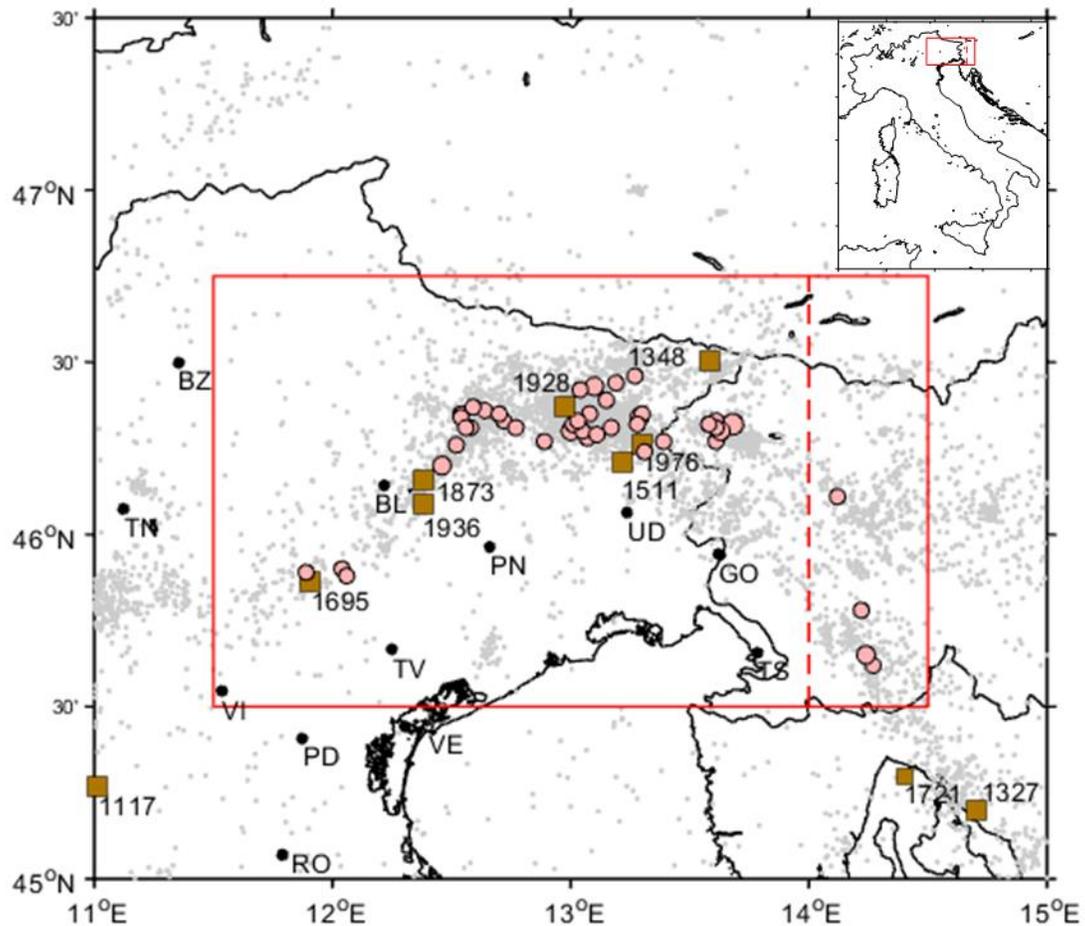

Fig. 1: Area of analysis. Red Rectangle: analyzed region. Dashed line: longitude 14. The inset shows the location of the study area. Circles: epicenters of analyzed clusters mainshocks (see Table 1). Brown squares: historical seismicity (earthquakes with intensity ≥IX). Grey dots: earthquakes recorded in the region from 1977. Main towns: TS=Trieste, GO=Gorizia, UD=Udine, PN=Pordenone, VE=Venice, TV=Treviso, BL=Belluno, PD=Padua, RO=Rovigo, VI=Vicenza, BZ=Bolzano, TN=Trento

The area they selected is between longitude 11.5 and 14. Most of the cells in the selected area have R>0.8. The analysis was performed on the whole catalogue because the applied cluster detection method analyses the catalogue as a whole and requires homogeneity in location accuracy.

In this paper, the changes in the characteristic of the catalogue in time do not affect the cluster identification procedure and the analysis can be extended also to less homogeneous parts of the catalogue. To use this feature, we applied the Kossobokov et al. 1999 scheme to the seismicity recorded after 2008, when the Antelope software was adopted with automatic data exchange with the Environmental Agency of the Republic of Slovenia (Agencija Republike Slovenije za Okolje - ARSO), the Austrian Central Institute for Meteorology and Geodynamics (Zentralanstalt für Meteorologie und Geodynamik - ZAMG) (Peruzza et al., 2015). For this seismicity, the quality of the catalogue improves eastwards, thus allowing the study of seismicity until longitude 14.5. In particular, before 2008 the ratio R in the analyzed area was less than 0.6, while in the last period it is over 0.9.

Thanks to this data integration, it was possible to include in the analysis the clusters in Knezak and in Postumia Slovenian regions after 2008.

Fig. 1 shows the region selected for the analysis between latitude 45.5 and 46.75, in good agreement with Peresan and Gentili (2018), extending the longitude to 14.5 after 2008.

## 2.2 Magnitude corrections and data completeness

The magnitude reported in the OGS catalogue is a duration magnitude. Due to problems connected with the method used to evaluate coda duration, the use of this magnitude causes some artifacts like e.g. the non-linearity of Gutenberg Richter relation (Gutenberg Richter, 1954) after year 1994 (Peresan and Gentili 2018). The problem is well known (see Bragato and Tento, 2005). Gentili et al. 2011 proposed a correction for magnitude smaller than 3.5 by using dataset with magnitude ≥ 1 as published by Bragato and Tento (2005) (see Eq. 1).

$$M_L = 1.377 * M_D - 1.376 \qquad (1)$$

Peresan and Gentili (2018) applied eq. 1 for earthquakes after 1994 and show that in this way the Gutenberg Richter relation is linear for $M_L \geq 1$. In this paper, we recomputed earthquakes magnitude using eq. 1 for magnitude ≤3.5. The equation is valid in the $M_L$ magnitude range [1, 3.5]. We therefore used the catalogue obtained after 1994 only for $M_L \geq 1$. According to Peresan and Gentili 2018, we made the conservative choice to fix completeness magnitude to $M_c=2.0$ until 1993, and to $M_c=1.5$ since 1994 in the whole analyzed area, if no other detail information is available.

## 2.3 Cluster identification

Several methods have been applied in the region for clusters identification. Gentili and Bressan 2008 manually selected a set of eight clusters with magnitude ≥3.7 in the region from 1977 to 2004, obtaining a relation between mainshock magnitude $M_m$ and duration $\tau$ and radius $\rho$ of the sequences in the area.

$$\tau = 10^{0.33 M_m + 0.42} \qquad (2)$$

$$\rho = 10^{0.41 M_m - 1} \qquad (3)$$

With the radius expressed in km and the duration in days.

Gentili (2010) declustered OGS catalogue from 1994 to 2004 by using a window-based method, which uses eq. (2) and (3) to detect the clusters: the earthquakes belonging to the cluster are the ones, and only the ones, within a radius ρ and a time interval τ from the mainshock. She compared the method with Reasenberg's declustering algorithm (Reasenberg, 1985) showing that window-based method outperforms the Reasenberg one in this area. Peresan and Gentili (2019) analyzed the clusters from 1994 to 2015 detecting them by a statistical approach called the ''nearest-neighbors'' method (NN) by Zaliapin and Ben-Zion (2013) in a sub-region characterized by quite homogeneous completeness magnitude and location accuracy. In addition, they compare characteristics of the clusters obtained by NN method with manual ones by Gentili and Bressan (2008) and with two widely used windows methods applying Gardner-Knopoff and Uhrhammer equations (Gardner and Knopoff, 1974; Uhrhammer, 1986). They show good performances of the NN method while the window-based method based on Gardner-Knopoff and Uhrhammer equation are not appropriate for the region. In this paper, we considered a larger area and longer time span than in Peresan and Gentili (2019). For this reason, the NN is not appropriate, because it implicitly assumes homogeneous location accuracy in statistical discrimination of clustered and background seismicity. In addition, the NN method is not designed as a real time application, because two unrelated events can be merged in the same cluster afterwards, if an earthquake is found in the middle.

For this reason, we applied a window-based method for clusters detection based on Gentili and Bressan (2008) regression. In particular, we used equation (2) for time duration, while we added 2 km to the radius ρ (epicentral distance):

$$\rho' = 10^{0.41 M_m - 1} + 2 \qquad (4)$$

in order to take into account the lower location accuracy in the first years of the catalogue, due to the lower number of seismic stations (for further details on the location uncertainty see Gentili et al. 2011).

Comparing the cluster lists obtained by using eq. (3) and (4), the clusters are similar, and the differences are relevant only before 1980, when location accuracy is poorer. For subsequent years, enlarging the space window either causes no change in the clusters or the sporadic addition of one earthquake with magnitude generally < 1 in some clusters.

Like in Gentili and Di Giovambattista (2017), we used an operational definition of clusters that defines ''mainshock'' as the first shock of the cluster over a given threshold magnitude. We adopted this criterion to be able to apply the procedure immediately after the occurrence of a shock, without waiting to verify if a stronger earthquake followed. For this reason, it may happen that the "mainshock" is not the largest shock of the cluster. To be able to classify clusters as type A or B, given a mainshock of magnitude $M_m$, we need at least one aftershock with magnitude $M \geq M_m - 1$ for type A, while for B clusters a completeness magnitude $M_c$ inside the cluster $M_c \geq M_m - 1$ is needed in order to exclude that the cluster could be of type A with large aftershocks missed.

We detected 50 earthquake clusters with mainshock $M_m \geq 3.7$ with more than one event from 1977 to the end of 2018 in OGS catalogue. The choice of a threshold magnitude of 3.7 is a compromise to maximize the number of available clusters, taking into account the completeness magnitude of the region (1.5-2). In the following analysis, eight clusters have been eliminated because they have longitude >14 and they occurred before 2008.

Besides the higher completeness magnitude, the main problem in handling these data is location and magnitude inaccuracy, due to bad station coverage of the regional OGS network. Table 1 shows the selected database of 42 clusters. Fig. 1 shows clusters mainshock location (circles).

| Place | Date | Time | ML | Lat | Lon | Naft | Mc | Ty | Dm | Dt |
|---|---|---|---|---|---|---|---|---|---|---|
| Trasaghis 77a | 03/07/1977 | 11:44:56 | 3.7 | 46.28 | 13.07 | 17 | 2.0 | B | 1.6 | 36.79 |
| Trasaghis 77 | 16/09/1977 | 23:48:07 | 5.2 | 46.30 | 13.00 | 483 | 1.2±0.2 | A | 1.0 | 11.08 |
| Pontebba 78 | 20/02/1978 | 12:13:34 | 4.0 | 46.46 | 13.27 | 8 | 2.0 | B | 2.1 | 28.83 |
| Gemona 78 | 03/04/1978 | 10:49:46 | 4.2 | 46.31 | 13.17 | 88 | 0.4±0.1 | B | 1.1 | 0.16 |
| Ampezzo 78 | 12/12/1978 | 15:14:49 | 4.4 | 46.33 | 12.72 | 10 | 2.0 | B | 1.8 | 25.62 |
| Chiusaforte 79 | 18/04/1979 | 15:19:19 | 4.8 | 46.34 | 13.29 | 366 | 0.9±0.1 | B | 1.1 | 61.78 |
| Chiusaforte 80 | 23/03/1980 | 16:56:36 | 3.7 | 46.35 | 13.30 | 26 | 2.0 | B | 1.3 | 0.47 |
| Lusevera 81 | 30/08/1981 | 23:30:29 | 3.9 | 46.32 | 13.28 | 26 | 2.0 | A | 0.9 | 0.22 |
| Uccea 83 | 10/02/1983 | 22:30:34 | 4.2 | 46.27 | 13.39 | 8 | 2.0 | B | 2.7 | 21.19 |
| Monte Pramaggiore 86 | 29/08/1986 | 14:57:01 | 4.0 | 46.35 | 12.54 | 1 | 2.0 | A | 0.1 | 0.00 |
| Mena 88 | 01/02/1988 | 14:21:38 | 4.1 | 46.35 | 13.08 | 257 | 1.1±0.1 | A | 0.3 | 3.22 |
| Pielungo 91 | 11/06/1991 | 08:05:54 | 3.9 | 46.27 | 12.89 | 9 | 2.0 | B | 1.2 | 0.03 |
| Tarcento 91 | 05/10/1991 | 05:14:58 | 3.8 | 46.24 | 13.31 | 51 | 2.0 | A | 0.7 | 0.40 |
| Claut 94 | 20/04/1994 | 21:25:26 | 3.7 | 46.34 | 12.54 | 28 | 1.5 | A | 0.7 | 1.25 |
| Claut 96 | 27/02/1996 | 11:13:46 | 3.8 | 46.31 | 12.58 | 43 | 1.5 | A | -0.5 | 46.07 |
| Kobarid 98 | 12/04/1998 | 10:55:33 | 5.6 | 46.32 | 13.68 | 733 | 1.4±0.1 | A | 1.0 | 23.66 |
| Trasaghis 98 | 28/05/1998 | 09:32:19 | 4.1 | 46.30 | 13.05 | 16 | 1.5 | B | 1.2 | 0.00 |
| Kobarid 99 | 13/05/1999 | 16:06:52 | 3.8 | 46.27 | 13.61 | 11 | 1.5 | A | 0.6 | 9.93 |
| Sernio 02 | 14/02/2002 | 03:18:03 | 4.9 | 46.43 | 13.10 | 45 | 1.5 | B | 1.9 | 11.32 |
| Forni di sotto 02 | 06/05/2002 | 03:24:17 | 3.8 | 46.36 | 12.64 | 2 | 1.5 | B | 1.2 | 2.44 |
| Kobarid 02 | 30/09/2002 | 02:48:32 | 3.9 | 46.33 | 13.61 | 12 | 1.5 | A | 0.9 | 0.01 |
| Sernio 03 | 11/03/2003 | 05:57:09 | 3.7 | 46.44 | 13.19 | 3 | 1.5 | B | 2.7 | 0.01 |
| Tramonti 03 | 30/08/2003 | 09:10:52 | 3.7 | 46.31 | 12.77 | 4 | 1.5 | B | 1.8 | 3.24 |
| Kobarid 04 | 12/07/2004 | 13:04:06 | 5.1 | 46.30 | 13.63 | 301 | 1.1±0.1 | B | 1.5 | 1.65 |
| Forni di Sotto 04 | 29/08/2004 | 00:04:42 | 3.8 | 46.35 | 12.70 | 2 | 1.5 | B | 1.5 | 0.04 |
| Claut 07 | 26/02/2007 | 05:50:46 | 3.9 | 46.26 | 12.52 | 3 | 1.5 | A | 0.1 | 0.35 |
| Trasaghis 08 | 29/02/2008 | 11:41:59 | 3.7 | 46.32 | 13.01 | 8 | 1.5 | B | 1.8 | 3.51 |
| Postumia 10 | 15/01/2010 | 14:20:54 | 4.0 | 45.78 | 14.22 | 30 | 1.5 | B | 1.7 | 49.72 |

| Place | Date | Time | ML | Lat | Lon | Naft | Mc | Ty | Dm | Dt |
|---|---|---|---|---|---|---|---|---|---|---|
| Knezak 10 | 15/09/2010 | 02:21:18 | 3.9 | 45.62 | 14.27 | 28 | 1.5 | A | 0.0 | 0.00 |
| Valdobbiadene 11 | 13/09/2011 | 18:35:24 | 3.7 | 45.90 | 12.04 | 11 | 1.5 | A | 0.5 | 0.01 |
| Alpago 12 | 09/06/2012 | 02:04:57 | 4.5 | 46.20 | 12.46 | 19 | 1.5 | B | 2.1 | 14.07 |
| Claut 13 | 12/02/2013 | 18:12:44 | 3.8 | 46.31 | 12.56 | 7 | 1.5 | B | 2.8 | 0.01 |
| Knezak 14 | 22/04/2014 | 08:58:28 | 4.7 | 45.65 | 14.24 | 32 | 1.5 | B | 2.2 | 0.44 |
| Moggio 15 | 30/01/2015 | 00:45:50 | 4.1 | 46.39 | 13.15 | 9 | 1.5 | B | 2.9 | 18.54 |
| Valdobbiadene 15 | 12/05/2015 | 02:02:50 | 3.7 | 45.88 | 12.06 | 26 | 1.5 | A | 0.1 | 3.15 |
| Cavaso del Tomba 15 | 18/08/2015 | 20:10:03 | 3.8 | 45.89 | 11.89 | 10 | 1.5 | B | 1.5 | 3.25 |
| Kobarid 15 | 29/08/2015 | 18:47:04 | 4.0 | 46.31 | 13.61 | 21 | 1.5 | B | 1.8 | 24.30 |
| Kobarid 18 | 17/01/2018 | 10:22:20 | 3.8 | 46.32 | 13.58 | 13 | 1.5 | B | 1.6 | 28.92 |
| Arta 18 | 19/01/2018 | 17:39:43 | 3.8 | 46.42 | 13.04 | 2 | 1.5 | B | 2.1 | 0.01 |
| Forni di Sopra 18 | 25/02/2018 | 08:16:30 | 3.9 | 46.37 | 12.59 | 54 | 1.5 | A | 0.2 | 0.32 |
| Gemona 18 | 09/05/2018 | 21:48:03 | 3.7 | 46.29 | 13.11 | 3 | 1.5 | B | 2.6 | 0.06 |
| Cavazzo 18 | 11/08/2018 | 03:30:39 | 3.9 | 46.33 | 13.03 | 23 | 1.5 | A | 0.9 | 0.02 |

Table 1: Analyzed clusters recorded by OGS catalogue. Place: Name of the closest town/village to the mainshock epicenter followed by the last two digits of the year. Date and Time: date and time of the mainshock in the format dd/mm/yyyy hh:mm:ss. ML=local magnitude of the mainshock. Lat and Lon: North Latitude and East Longitude of the mainshock (degrees): Naft: total number of recorded aftershocks. $M_c$: completeness magnitude. Ty: cluster typology (A or B). Dm and Dt: difference in magnitude and time (in days) between the mainshock and the strongest aftershock.

The completeness magnitude changes in space and time inside the selected area of analysis (Gentili et al. 2011). In addition, clusters have generally been recorded with higher accuracy with respect to the background seismicity, thanks to the installation of temporary seismic networks during relevant clusters. For this reason, if at least 80 aftershocks were recorded, we calculated the completeness magnitude $M_c$ for each cluster independently. The completeness magnitude was evaluated by the maximum curvature method and the uncertainty by bootstrapping. For a discussion on the choice of completeness magnitude definition, see Gentili and Di Giovambattista (2017). The obtained completeness magnitudes are comparable with the results of Gentili et al., 2011. For clusters with less than 80 aftershocks, according to Peresan and Gentili 2018, the completeness magnitude has been fixed to $M_c$=2.0 until 1993, and to $M_c$= 1.5 since 1994. It may be argued that a small number of earthquakes composes some of the selected clusters. However, the topic of this paper is not a cluster characteristics analysis that requires tens or hundreds of aftershocks, but vice-versa, an analysis in the space-time region where the cluster is expected, in order to forecast the occurrence of a strong following earthquake. A small number of aftershocks in the analyzed space-time area in the first hours/days after the mainshock is itself an important feature for strong following earthquake forecasting.

## 3. The pattern recognition approach

We used the method for strong forthcoming earthquake estimation approach developed in Gentili and Di Giovambattista (2017). The main idea is to develop an adaptive software package (here and after NESTORE - NExt STrOng Related Earthquake) analyzing the seismic data at increasing time intervals [0, T] after the mainshock and able to adapt itself to the characteristics of the region and to available data. NESTORE will analyze a pre-defined kernel of seismicity features (e.g. number of earthquakes recorded, their energy and distribution etc.; see Vorobieva 1999, Vorobieva and Panza, 1993, Kossobokov et al., 1999, Di Giovambattista and Tyupkin, 2002, Gentili and Bressan 2008), tested as being successful in different regions, and a set of other features not available in all of the regions (e.g. focal mechanism of the mainshock, accurate estimation of mainshock depth) and that can be used only in some regions.

In Gentili and Di Giovambattista (2017), e.g., we used a set of features based on the subsequent earthquake, that can be applied also to the OGS catalogue, and a set of mainshock or foreshock based features that cannot be applied in this case due to unavailability or unreliability of data for smaller or older earthquakes or due to the small number of foreshocks in the region.

Our analysis concentrates on features based on earthquakes following the mainshock proposed in Gentili and Di Giovambattista (2017). For a detailed list of references on where the features were proposed for the first time in literature, see Gentili and Di Giovambattista, 2017. The hypothesis for predicting strong aftershocks, based on the adopted features, is the same of Vorobieva (1999): before strong mainshocks, a change in earthquakes flow, which becomes more intense and irregular in space and time, has been observed ( Keilis-Borok and Rotwain, 1990 , Keilis-Borok and Kossobokov, 1990) and interpreted as a symptom of instability of a nonlinear system corresponding to earthquake-generating faults; similar symptoms after a mainshock could precede the occurrence of a second large event. In order to obtain a good statistical significance for all the features based on the subsequent earthquake, NESTORE algorithm needs a completeness magnitude $M_c$ for the cluster of $M_m$-3, where $M_m$ is the mainshock magnitude.

However, the subset of clusters in the OGS catalogue satisfying this condition (here and after database D3) is very small and consists of six B clusters and three A clusters. For this reason, we tested separately the performances of the features needing a completeness magnitude of $M_m$-2 on a larger database (database D2 including nine A and twenty three B clusters) and we developed two different NESTORE instances, called NESTORE_M2 and NESTORE_M3, to be applied depending on the difference $M_m$-$M_c$.

For each feature, pattern recognition is done in two steps: a training set is defined, composed by patterns (vectors of one feature in this case) corresponding to clusters of known class (A or B), and, after the training, if successful, the classifier is able to label new clusters as A or B based on the corresponding feature value. This training procedure, where the class of the patterns of the training set is known, is called "supervised".

### 3.1 Analyzed features

Only earthquakes with magnitude greater or equal to the mainshock magnitude ($M_m$) minus a given feature-dependent threshold *ThM* are considered. Each function is calculated at time intervals [$s_1$,$s_2$], where $s_1$ is feature dependent and $s_2$ increases to simulate the increasing information with time passing after a strong earthquake.

The features analysed are the following:

1. *N*: number of events with $M \geq M_m$-3.

2. N2: number of events greater than $M_m$-2. Due to higher aftershocks magnitude respect to N, less problems connected with completeness magnitude arise immediately after the mainshock and a smaller $s_1$ can be selected. It may happen that N2>N.

3. *S*: normalized aftershocks' source area;

$$S(i) = \sum_i 10^{(m_i - M_m)} \tag{5}$$

where $m_i$ is the magnitude of the $i^{th}$ event in the selected time interval.

4. $V_m$: cumulative variation of magnitude from event to event;

$$V_m(i) = \sum_i |m_i - m_{i-1}| \tag{6}$$

5. $V_{med}$: cumulative variation of average magnitude from day to day (starting one hour after the mainshock ):

$$V_{med}(i) = \sum_i |\mu_i - \mu_{i-1}| \qquad (7)$$

where $\mu_i$ is the average magnitude of the mainshocks on the $i^{th}$ day.

6. $V_n$: cumulative variation of the number of aftershocks from day to day (starting one hour after the mainshock):

$$V_n(i) = \sum_i |n_i - n_{i-1}| \qquad (8)$$

where $n_i$ is the number of aftershocks that occurred on the $i^{th}$ day ($M \geq M_m-3$).

7. $Z$: linear concentration of aftershocks

$$Z(i) = \frac{mean(10^{0.69 m_i - 3.22})}{mean(r_{ij})} \qquad (9)$$

where $r_{ij}$ is the distance between the generic $i^{th}$ and $j^{th}$ aftershock.

8. *SLCum*: cumulative deviation of S from the long-term trend on increasing length windows;

$$SLCum(i) = \sum_i abs[S(t_i) - S(t_{i-1}) \frac{i \cdot dt}{(i-1) \cdot dt}] \qquad (10)$$

Where $t_i = s_1 + i \cdot dt$, $S(t_i)$ is S calculated on the time interval $[s_1, t_i]$ and $dt$=6 hour.

9. *SLCum2*: cumulative deviation of *S* from the long-term trend on sliding windows;

$$SLCum2(i) = \sum_i abs[S([s_1 + (i-1) \cdot dt, s_1 + i \cdot dt]) - S([s_1 + (i-1) \cdot dt, s_1 + (i-1) \cdot dt + d\tau]) \frac{dt}{d\tau}] \qquad (11)$$

where $S[a, b]$ is *S* calculated over the generic time interval $[a, b]$. Differently from *SLCum*, the window does not start at a fixed time close to the mainshock origin time.

10. *Q*: Normalized Radiated Energy;

$$Q(i) = \frac{\sum_i E_i}{E_m} \qquad (12)$$

where $E_m$ is the energy of the mainshock and $E_i$ is the energy of the $i^{th}$ aftershock. The energy E (in Joules) of an event of magnitude M is obtained by the equation $Log_{10}(E) = \frac{3}{2}M + 4.8$ (Gutenberg and Richter, 1956).

11. *QLCum*: cumulative deviation of *Q* from a long-term trend on increasing length windows;

$$QLCum(i) = \sum_i abs[Q(t_i) - Q(t_{i-1}) \frac{i \cdot dt}{(i-1) \cdot dt}] \quad (13)$$

where $t_i = s_1 + i \cdot dt$, and $Q(t_i)$ is calculated on the time interval $[s_1, t_i]$.

12. *QLCum2*: cumulative deviation of *Q* from the long-term trend on sliding windows;

$$QLCum2(i) = \sum_i abs[Q([s_1 + (i-1) \cdot dt, s_1 + i \cdot dt]) - Q([s_1 + (i-1) \cdot dt, s_1 + (i-1) \cdot dt + d\tau]) \frac{dt}{d\tau}] \quad (14)$$

where $Q[a, b]$ is *Q* calculated on the generic interval $[a, b]$.

Table 2 shows the values of $s_1$ and *ThM* adopted for each feature.

| Feature | $s_1$ | ThM |
|---|---|---|
| N | 1 hour | 3 |
| N2 | 1 s | 2 |
| S | 1 hour | 2 |
| $V_m$ | 1 hour | 3 |
| $V_{med}$ | 3 days | 3 |
| $V_n$ | 3 days | 3 |
| Z | 1 hour | 2 |
| SLCum | 12 hours | 2 |
| SLCum2 | 12 hours | 2 |
| Q | 1 hour | 2 |
| *QLCum* | 12 hours | 2 |
| *QLCum*2 | 12 hours | 2 |

Table 2: Values of $s_1$ and *ThM* for the analyzed features.

According to Gentili and Di Giovambattista (2017), each feature has been evaluated by a pattern recognition approach using an independent decision tree (e.g., Breiman et al., 1984; Jang et al., 1997). In particular, for each feature, a one-node decision tree is evaluated; in other words, the algorithm identifies for each feature a threshold *Th* so that (most of) the clusters in A class have feature value greater than *Th*, while (most of) the clusters of type B have feature value smaller than *Th*.

In order to test the performances, we applied the Leave-One-Out (or LOO) method because of its limited computational cost, the small datasets used, and because it allows enlarging the training set as much as possible to capture the variability of the features inside the same class. The LOO method trains the classifier by all the available patterns except one (training set) and tests the performances on the pattern which have not been used in the training (test set). The procedure is repeated for all patterns. For details see Gentili and Di Giovambattista, 2017. The NESTORE method estimates the performances of each feature in time. Even if the analysis is done a-posteriori, it simulates the performances in real case as after a strong earthquake the classification can be done using data only until the current time. Performance estimate is made in terms of Precision, Recall, Accuracy and Informedness (see Appendix for details). The first three performances evaluators are defined between 0 and 1, where 0 means the worst performances and 1 the best ones, the last one ranges between -1 and 1, -1 being the worst performances and 1 the best. The typical trend of these performance evaluators in NESTORE applications is an increase on the evaluator value as the observation time increases, until a maximum, followed by a constant value or a decrease if the observation time is longer. The method therefore defines a time $S_2$ corresponding to the best performance of the feature and, when longer observation times are available, uses the feature value at $S_2$ time.

## 3.2 Combining classifiers

NESTORE method trains a set of classifiers based on independent features. The different classification results need to be combined in a unique classification that may be binary ("Class A" or "Class B") or continuous ("Probability of Class A"). The simplest approach to classification combination is a voting procedure: each classifier has one vote (A or B) and if the number of A ($N_A$) is greater than the number of B ($N_B$), the classification is A, otherwise B. Some easy amendments to this method are (1) The final classification is A only if $N_A >= N_B+D$, where D is a natural number >0 (e.g. Vorobieva and Panza (1993), (2) The probability to be in the A class is $N_A/(N_A+N_B)$. The previous approaches imply that all classifiers are reliable in the same way. This is generally not true in real cases. In Gentili and Di Giovambattista (2017), we proposed a different method, according to which the Informedness of each feature is considered. Informedness is a measure of how informed the classifier about the class is (see Appendix); more reliable classifiers have higher Informedness. Gentili and Di Giovambattista (2017) used the Informedness of each feature to weight the vote of the corresponding classifier, so that more reliable features have higher score in the voting procedure. This approach takes into account each classifier by using only one score that summarizes the classifier's performances. However, we noticed that some features supply very interesting information when they are over the selected threshold *Th*, while under the threshold the results are more confused, or vice-versa. For example, in our database, if 6 hours after the mainshock, $Z \geq 0.018$, all the clusters are of type A (very reliable information), while, vice-versa, if $Z < 0.018$ the 22% of the clusters are of type A and the 78% of type B (less reliable information). In this new version of the algorithm, we estimated for each time step and each feature the probability *p* to have an A cluster over and under the corresponding threshold *Th*. In the case of a new cluster arrival, comparing each feature value with its corresponding threshold, we had N estimates of the probability that it is an A cluster (strong following earthquake is expected), where N is the number of features. We estimated the overall probability by using a Bayesian approach in combining independent classifiers.

According to e.g. Bailer-Jones et al. 2011, the posterior probability that class is C using independent information (feature) $D_1 \ldots D_n$ is:

$$P(C|D_1 \ldots D_N) = a \frac{\prod_{n=1}^{N} P(C|D_n)}{P(C)^{\{N-1\}}} \tag{15}$$

Where $P(C)$ is the probability to have class $C$, $P(C|D_n)$ is the posterior probability that class is C given $D_n$, and $a$ is a normalization factor such that $\sum_k P(C_k|D_1, ... D_N) = 1$, where $C_k$ are the classes of the problem. In this case, we had two classes: $A$ and $B$.

For each time step, assuming

$$P(A) = \frac{N(A)}{N(A)+N(B)} \tag{16}$$

and

$$P(B) = \frac{N(B)}{N(A)+N(B)} \tag{17}$$

where N(A) and N(B) are the number of A and B clusters in the database, respectively, we obtained

$$P(A|D_1 ... D_N) = \frac{[N(B)]^{N-1} \prod_{n=1}^{N} p_n}{[N(B)]^{N-1} \prod_{n=1}^{N} p_n + [N(A)]^{N-1} \prod_{n=1}^{N} (1-p_n)} \tag{18}$$

Where $p_n=P(A|D_n)$ is the probability to have a type A cluster given a value $D_n$ of the $n$ feature. We evaluated $p_n$ from the training set as the percentage of A clusters (divided by 100) over or under the corresponding threshold.

One advantage of using this approach is that it takes into account also the number of A and B clusters in the database that is very important for unbalanced classes like in our case (the B clusters are approximately the double of the A ones).

## 4. Results

### 4.1 Clusters statistical analysis

In order to evaluate the magnitude characteristics of the stronger aftershocks of the clusters in the region, it is necessary to guarantee that the network was able to record it. Since we used 0.1 accuracy in magnitude estimation, we checked in the database of Table 1 if the strongest aftershock recorded magnitude has at least magnitude $M_c$-0.1. While all the sixteen A clusters satisfy this condition, five B clusters (here and after BN clusters) do not satisfy it. BN clusters have magnitude of the mainshock between 3.7 and 4.2 and should be carefully included in the analysis. In fact, *Dm* of BN clusters could be overestimated, because an aftershock of magnitude greater than the recorded one, but lower than the completeness magnitude, could be lost by the seismic network. For these clusters *Dm* listed in Table 1 should be regarded as an upper limit for *Dm*. However, there is also a lower limit for *Dm*, because the strongest aftershock that could be missed has maximum magnitude $M_c$-0.1. Therefore, for these clusters we do not have information on the value of *Dm*, but a range of possible values. In particular, BN clusters have *Dm* between 2.2 and 2.9.

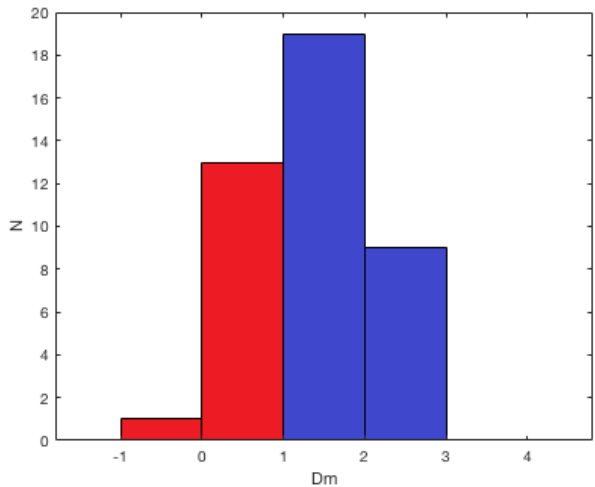

Fig. 2: Distribution of Dm. Red: A class; Blue: B class.

Figure 2 shows the distribution of *Dm* of Table 1. Negative values of *Dm* correspond to the cases in which a strongest event followed the "mainshock" automatically selected by the algorithm. We have only one case, the Claut 96 swarm, in which the strongest 4.3 mainshock followed the first one, of magnitude 3.8, 46 days later. If we considered the strongest event as mainshock, the value of *Dm* for Claut cluster was 0.8, due to a ML=3.5 earthquake a month later. BN clusters are considered in the histogram of Fig. 1 in the bar between 2 and 3. The histogram is approximately symmetric around the interval *Dm*= [1, 2] in good agreement with Båth law (Båth, 1965). In addition, the histogram clearly states the prevalence of B clusters (in blue): about the double of the A ones (in red). The mean value of *Dm* ranges between 1.3 and 1.4, considering the uncertainty due to BN clusters and the uncertainty of the mainshock determination in Claut 96 swarm, corresponding to a mean of Dm equal to 0.5 for the A class and a mean of Dm between 1.7 and 1.8 for B class.

Fig. 3 shows the distribution of the mean of *Dm* as function of the mainshock magnitude. The green bars correspond to the distribution if for the strongest aftershock magnitude we take the strongest recorded aftershock magnitude; the grey bars correspond to the distribution if we substitute the recorded magnitude of the aftershocks of the five BN clusters with $M_a=M_c-0.1$, corresponding to the strongest magnitude of an event that can be lost by the network. Green and grey bars therefore respectively represent the higher and the lower limits of the possible value of *Dm* value compatible with our information.

Data with magnitude greater than 4.5 are coincident because the BN clusters have mainshock magnitude <4.5. For $M_m$>4.5 all histograms show a general decrease of *Dm* (increase of strongest aftershock magnitude) as *Mm* increases, while for $M_m$<4.5 *Dm* increases as *Mm* increases. Therefore, at least in the studied region, different trends can be found for small mainshock and large mainshock clusters: while for magnitude <4.5 smaller mainshock clusters are the ones with larger aftershocks compared to the mainshock (smaller *Dm*), for $M_m$>4.5 the trend is the opposite and, as the mainshock magnitude increases, also the aftershock magnitude increases with respect to the mainshock, in general agreement with the hypothesis of Gentili and Di Giovambattista (2017) that larger mainshocks correspond, in general, to A clusters. It is important to remark, however, that most of the clusters of the database have $M_m$<5 so these data should be confirmed by further investigations.

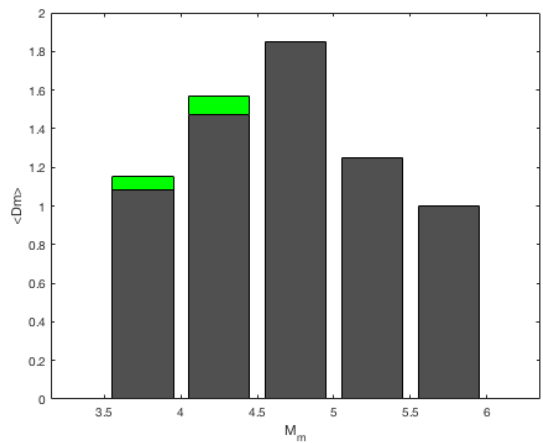

Fig. 3: mean of *Dm* as function of the mainshock magnitude

After a strong earthquake, two are the main questions important for population and the civil protection point of view: the first one is "Will other strong earthquakes occur?" and we tried a statistical answer in Fig. 2 and 3. The second question is "When will it be sure enough to come back home and start reconstruction?". This is a very important question, because a too early comeback in the buildings has caused a large toll of deaths in Friuli in September 1976 (four months later the 6 May 1976 stronger earthquake) and, more recently, during Umbria-Marche cluster, in September 1997, during the inspection of the Assisi cathedral, few hours after a strong foreshock. Even if we cannot give an exact prevision for the single cluster, we analyze the problem from the statistical point of view in the following. In this analysis, we need further information than in the previous one, because we do not only need to know the magnitude of the strongest aftershock, but also when it was. For this reason, we need to reduce our dataset to the only clusters whose strongest aftershock magnitude is greater or equal to the completeness one. The resulting database is composed by 36 clusters: 16 A clusters and 20 B ones.

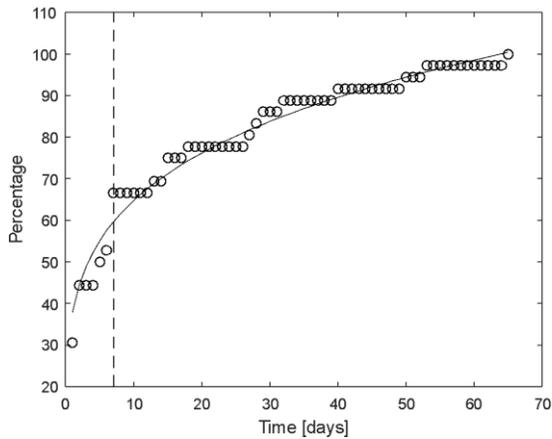
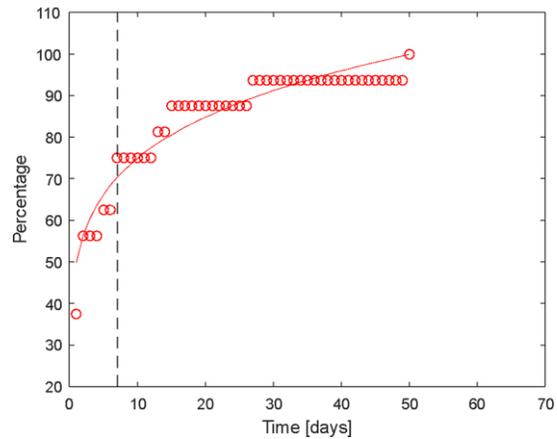

(a)  (b)

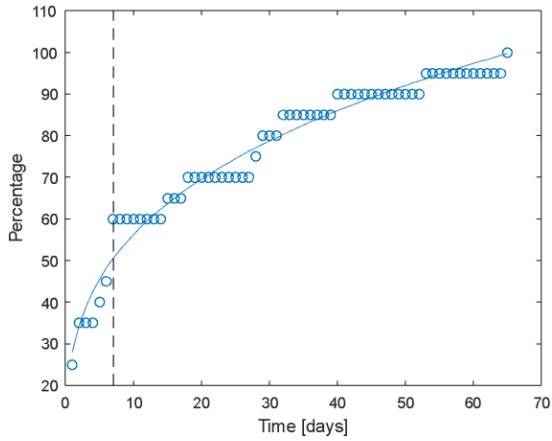
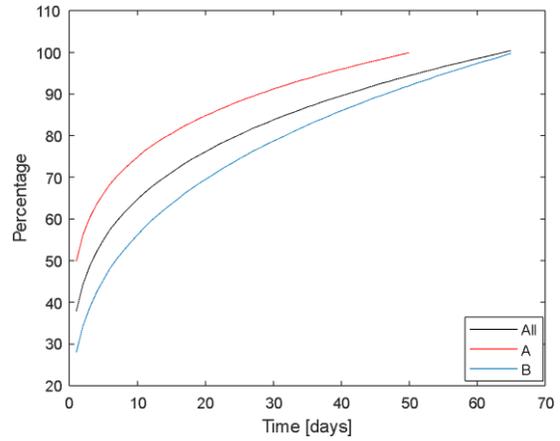

(c)                                            (d)

Fig. 4: Circles: percentage of clusters that already had the strongest aftershock in time. Continuous lines: fit of the data accordingly with eq. (19). Dashed line: 7 days after the mainshock. (a) All data (b) Type A clusters data (c) Type B clusters data (d) Comparison between the previous fit

Considering all the clusters, the strongest aftershock is in the first 6 hours after the mainshock in the 31% of the cases, within the first day in the 44% of the cases, and within 4 days after the mainshock in the 67% of the cases.

As the most dangerous clusters are type A, due to strong aftershock magnitude similar to the one of the mainshock, the higher danger is in the first hours and days after the mainshock. On the other side, the first hours are crucial to forecast whether the analyzed cluster is an A or a B cluster.

Figure 4 shows the percentage of clusters which already had the stronger aftershock at a given time T after the mainshock, vs time T. The percentage of clusters increases fast within the first 7 days, while the trend becomes smoother for longer time intervals.

In order to understand if there is some mathematical description of the trend of the percentage (Perc) as function of T, the curves have been fitted for T≥ 6 hours for all data, with the equation:

$$\text{Perc} = \lambda T^{\gamma} \tag{19}$$

The values of $\gamma$ and $\lambda$ are listed in Table 3 for the three curves. The fit supplies good results for T≥12 hours, while it overestimates the percentage for shorter times.

|         | γ           | λ         |
|---------|-------------|-----------|
| All data | 0.234±0.003 | 37.8±0.3 |
| A data  | 0.178±0.005 | 49.7±0.7 |
| B data  | 0.306±0.004 | 27.9±0.6 |

Table 3: Parameters of eq. 19

The A clusters show a smoother trend than the B clusters and reach the 100% of clusters before B ones only because after 1 day 56% of A cluster already had the strongest aftershock, while only 35% of B clusters did.

For all the analyzed clusters, the strongest aftershock has been recorded within 70 days from the mainshock. Taking in mind the 1976 Friuli cluster, not contained in this database and stronger than the analyzed clusters, with destructive earthquakes 4 months after the first aftershock, we investigated if there is a dependence of $Dt$ on the mainshock magnitude. In Figure 5, we show the mean $Dt$ in dependence of the mainshock magnitude. For magnitude <5, the value of $Dt$ increases with the mainshock magnitude. For larger magnitudes, data are less reliable, due to the small number of clusters involved. While the cluster with magnitude ≥5.5 shows mean $Dt$ larger than clusters with mainshock of magnitude <5, a decrease of <$Dt$> is shown between 5 and 5.5. A larger database is necessary in order to verify if the hypothesis is correct.

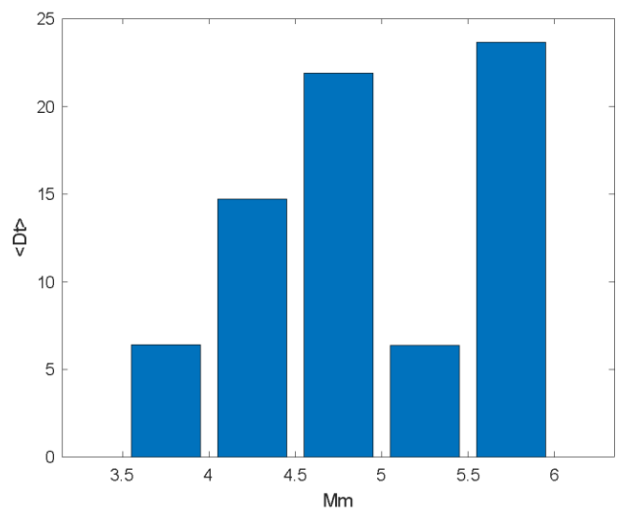

Fig 5: Mean of $Dt$ as function of the mainshock magnitude

## 4.2 Analysis by NESTORE method

We calculated the values of the features described in section 4.1 for each cluster at different time steps, corresponding to enlarging time intervals after the mainshock. For each interval T, we calculated the features only using the data within the time interval, and we applied the LOO method in order to calculate the confusion matrix. This is done in order to simulate the sequences as if they were in real time. The ongoing seismic activity after the mainshock adds some information as T increases and therefore the features performances change in time. In particular, we analyzed the data using intervals [0, T] ending at T 6, 12, 18 hours and 1, 2, 3 days after the mainshock. Some features of the NESTORE method ($N$, $V_m$, $V_{med}$, $V_n$) require completeness magnitude $M_c \leq M_m$-3. This choice reduces the applicability of the whole method to nine clusters (database D3) of the initial 42, of which three type A and six type B clusters. This leads to instability of results, because when one A cluster is tested by the LOO method, the training set contains only two A clusters. The other features need a completeness magnitude $M_m$-2 in order to be evaluated; we selected a larger dataset (here and after D2), composed by all the clusters with $M_c \leq M_m$-2, instead of the dataset D3, for validation procedure. Dataset D2 is composed by 32 clusters, of which nine A type and twenty-three B type clusters, and contains database D3. The results we obtained on features validated on dataset D2 are therefore more reliable than the ones validated on D3. Like in Gentili and Di Giovambattista (2017) we selected a maximum time $S_2$ for feature evaluation that corresponds to the time where the performances are best in terms of Precision, Recall, Accuracy and Informedness.

Like in that paper, the condition for the feature to be reliable are:

1. The Accuracy is greater or equal to the one we can obtain by a constant response corresponding to the most populated class (B class).
2. The Informedness is greater than 0.

Both SLcum2 and $V_{med}$ have negative Informedness, while the Accuracy is smaller than the constant response; for this reason, they were not used in the NESTORE classification phase.

Fig. 6 displays the ROC diagram showing features performances for different time steps after the mainshock. The features shown for each interval [0, T] are the ones with $s_1 \leq T \leq S_2$. We used the database $D_2$ when possible, because the results obtained by the database D3 are sometimes unstable due to the small database involved. Not all of the features could be calculated for shorter times after the mainshock (see Table 3). The earlier classification, and therefore the earlier potential alarm of the method, can be produced 6 hours after the mainshock, using the features N, N2, S, Q and $V_m$, which supply good performances with a true positive rate between 0.56 and 0.67 and a false positive rate between 0.08 and 0.33, well over the line corresponding to random response (black dashed line). Z feature performances are over the black dashed line too, but the true positive rate (hit rate) is very small. By increasing time T other features become available and the performances of some features are improved. The best performances in terms of distance from the upper left corner in ROC diagram, are the ones of $V_m$ and N2 after 6 hours, N after 1 day, and $V_n$ and QLcum after three days (Informedness in the range [0.4, 0.7]. Unsurprisingly, SLcum2 and $V_{med}$ have poor performances, below the random response.

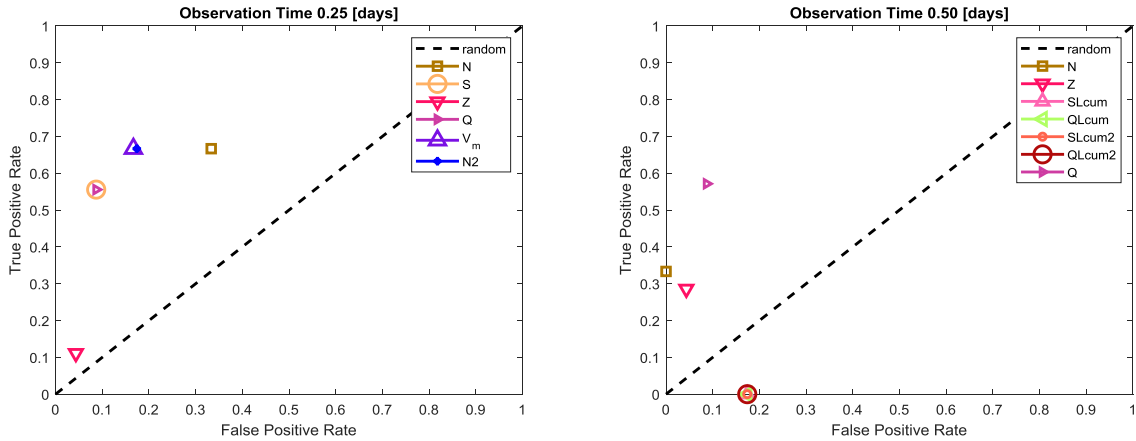

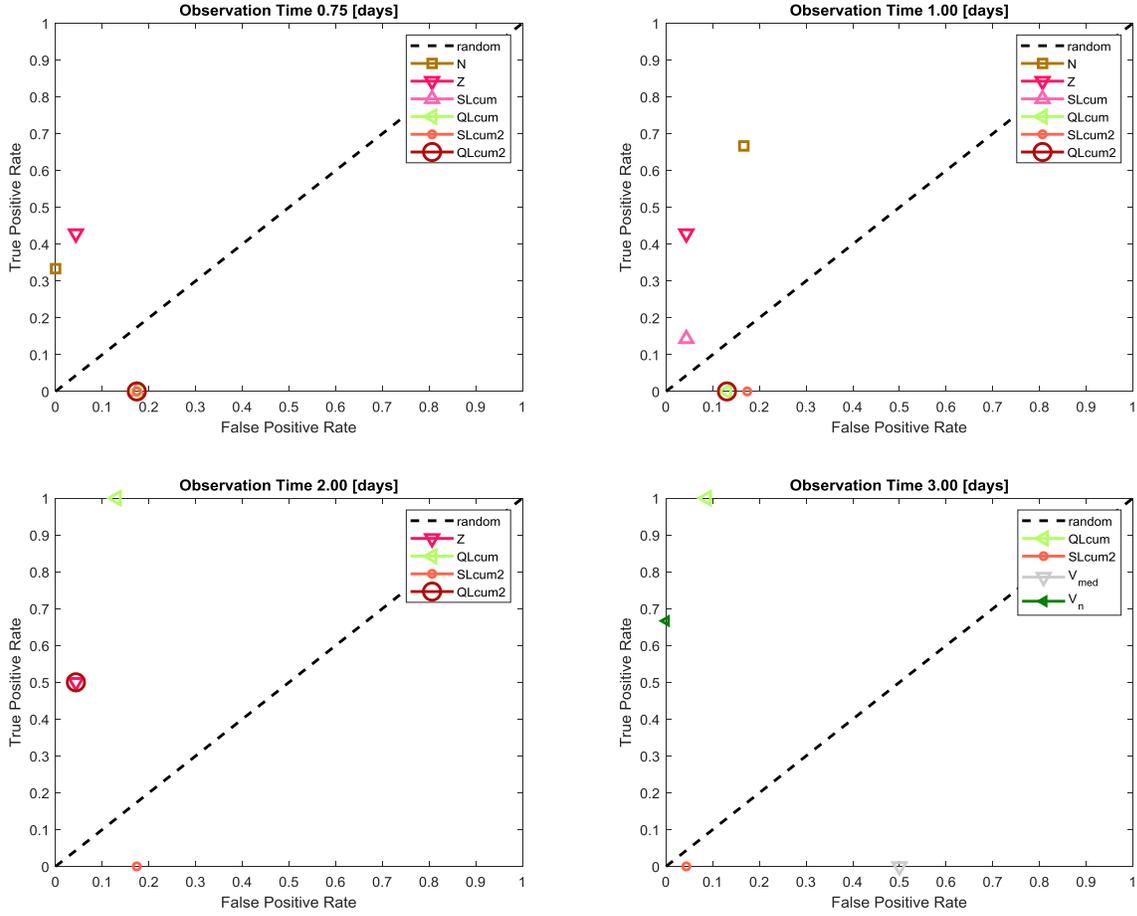

Fig. 6: ROC diagram showing the features performances at different observation times T.

We stop our analysis to T=3 days, because database D3 contains only two A clusters for larger T.

In order to develop a classifier that can be used for future forecasting, we used the whole databases available (D3 or D2 depending on the feature) to estimate the threshold for each feature. The obtained results should be tested on independent databases, e.g., in future analysis of future clusters. Table 4 shows the thresholds *Th* obtained by NESTORE algorithm using the D3 database for N, Vm and Vn features and D2 for all the other features. Thresholds are shown for Informedness>0, corresponding to performances over the random response line in ROC diagrams, and T≤$S_2$.

| Feature | $S_2$ | 6h | 12h | 18h | 1d | 2d | 3d |
|---|---|---|---|---|---|---|---|
| N | 1d | 2.5 | 6 | 11 | 11 | | |
| N2 | 6h | 0.5 | | | | | |
| S | 6h | 0.006 | | | | | |
| Z | 2d | 0.018 | 0.0165 | 0.0145 | 0.015 | 0.0155 | |
| SLCum | 1d | | -- | -- | 0.391 | | |
| QLCum | 3d | | -- | -- | -- | 3.6 | 4.2 |
| QLCum2 | 2d | | -- | -- | -- | 3.7 | |
| Q | 12h | 0.0005 | 0.001 | | | | |
| Vm | 6h | 0.1 | | | | | |
| Vn | 3d | | | | | | 8 |

Table 4. Thresholds evaluated by the algorithm on all available datasets for different time intervals (6h=6 hours, 1d=1 day etc.). "- -": the feature can be evaluated, but the Informedness is lower than zero and the feature is not used in classification.

Table 5 shows the probability for $P_L$ and $P_G$ to have an A cluster if the feature is lower or greater to the selected threshold *Th*. These probabilities can be substituted in eq (18) in place of $p_n$ to estimate the overall Bayesian probability to be in the A class. Fluctuations in probabilities for different databases are due to low number of data in D3 database. Changes in probabilities for $T>S_2$ are due to changes in the number of A clusters available increasing *T* (clusters with $Dt<T$ have been eliminated from the database). We noticed that $P_G$ coincides with the value of the precision for the feature estimated on the whole dataset.

| Feature | Data | 6h | | 12h | | 18h | | 1d | | 2d | | 3d | |
|---|---|---|---|---|---|---|---|---|---|---|---|---|---|
| | | $P_L$ | $P_G$ | $P_L$ | $P_G$ | $P_L$ | $P_G$ | $P_L$ | $P_G$ | $P_L$ | $P_G$ | $P_L$ | $P_G$ |
| N | D3 | 0.00 | 0.75 | 0.14 | 1.00 | 0.14 | 1.00 | 0.14 | 1.00 | 0.14 | 1.00 | 0.14 | 1.00 |
| N2 | D3 | 0.17 | 0.67 | 0.17 | 0.67 | 0.17 | 0.67 | 0.00 | 0.75 | 0.00 | 0.75 | 0.00 | 0.75 |
| | D2 | 0.13 | 0.60 | 0.13 | 0.50 | 0.13 | 0.50 | 0.13 | 0.50 | 0.09 | 0.50 | 0.09 | 0.50 |
| | D3 | 0.17 | 0.67 | 0.17 | 0.67 | 0.17 | 0.67 | 0.17 | 0.67 | 0.17 | 0.67 | 0.17 | 0.67 |
| S | D2 | 0.15 | 0.71 | 0.12 | 0.67 | 0.12 | 0.67 | 0.12 | 0.67 | 0.12 | 0.60 | 0.12 | 0.60 |
| Z | D3 | 0.25 | 1.00 | 0.14 | 1.00 | 0.14 | 1.00 | 0.14 | 1.00 | 0.14 | 1.00 | 0.14 | 1.00 |
| | D2 | 0.22 | 1.00 | 0.14 | 1.00 | 0.14 | 1.00 | 0.14 | 1.00 | 0.12 | 1.00 | 0.12 | 1.00 |
| SLCum | D3 | | | -- | -- | -- | -- | 0.33 | - | 0.33 | - | 0.33 | - |
| | D2 | | | -- | -- | -- | -- | 0.17 | 1.00 | 0.17 | 1.00 | 0.17 | 1.00 |
| QLCum | D3 | | | -- | -- | -- | -- | -- | -- | 0.00 | 0.75 | 0.00 | 0.75 |
| | D2 | | | -- | -- | -- | -- | -- | -- | 0.08 | 0.67 | 0.08 | 0.67 |
| QLCum2 | D3 | | | -- | -- | -- | -- | -- | -- | 0.14 | 1.00 | 0.14 | 1.00 |
| | D2 | | | -- | -- | -- | -- | -- | -- | 0.12 | 0.75 | 0.12 | 0.75 |
| Q | D3 | 0.17 | 0.67 | 0.17 | 0.67 | 0.17 | 0.67 | 0.17 | 0.67 | 0.17 | 0.67 | 0.17 | 0.67 |
| | D2 | 0.15 | 0.71 | 0.12 | 0.67 | 0.12 | 0.67 | 0.12 | 0.67 | 0.12 | 0.60 | 0.12 | 0.60 |
| Vm | D3 | 0.00 | 0.75 | 0.00 | 0.75 | 0.00 | 0.75 | 0.00 | 0.75 | 0.00 | 0.75 | 0.00 | 0.75 |
| Vn | D3 | | | | | | | | | | | 0.14 | 1.00 |

Table 5. Probability to have an A classification if the feature is over or under a given threshold for different time intervals (6h=6 hours, 1d=1 day etc.). $P_L$: probability to have an A cluster if the feature is lower than $Th$; $P_G$: probability to have an A cluster if the feature is higher or equal to $Th$; "- -": the feature can be evaluated, but the Informedness is lower than zero and the feature is not used in classification; "-": The probability cannot be evaluated for D3 dataset, because no data are available in that range so the feature is not used.

In order to have an estimate of the consistency of data with the selected thresholds and to outline clusters with an anomalous behavior, we applied NESTORE_M2 and NESTORE_M3 to the databases D2 and D3 respectively. We remark that (unlike the results presented in Figs. 6, 7) this is not an estimate of the method performances, because the training test and the test set coincide, and the performances could be overestimated due to data leakage (Kaufman et al, 2011).

When NESTORE_M2 applied to D2, $N$, $V_m$ and $V_n$ features cannot be used.

While 82% of the clusters are correctly classified, independently from the observation time T, there are 6 clusters (Trasaghis 77, Gemona 78, Trasaghis 98, Kobarid 99, Valdobbiadene 15, Forni di Sopra 18) supplying wrong performances. Of these clusters, Trasaghis 77 one supplies correct classification after one day, while the other clusters supply wrong classification for most features for all the time intervals T. In the discussion session we will make some hypothesis on why these clusters seem to belong to a different population.

When we applied NESTORE_M3 to the D3 database, $N$, $V_m$ and $V_n$ features. The overall performances are improved: Gemona78, misclassified for NESTORE_M2, is correctly classified by the NESTORE_M3, supplying more accurate results. No outliers are detected in D3 database, except Trasaghis 77 in the first day after the mainshock.

**4.2.1 Could 1976 September earthquakes be forecasted? Tests on independent data.**

In order to verify the algorithm performances on independent data, we tested NESTORE algorithm on three clusters not belonging to the database. For the selected time intervals for which NESTORE was trained ([0,0.25], [0,0.5], …. [0,3] see section 4.2) we extracted the features, compared them with the corresponding thresholds of Table 4, and applied equation (18) by using the probability values on Table 5.

The higher mainshock magnitude cluster we analyzed was the one occurred in Friuli in 1976. Data are not available in the OGS catalogue, because the OGS seismic network was installed one year after, so we used the CPTI15 earthquake catalogue (Rovida et al. 2011). The 1976 cluster mainshock had magnitude $M_L$= 6.5 and was on May 6, 1976; analysis starts from the mainshock, because the only foreshock (magnitude 4.6) was only one minute before the mainshock and could be easily identified as foreshock. The mainshock was followed by several earthquakes with magnitude between 4 and 5.1 during May; the seismicity decreased in the following three months, but in September a new swarm of seismicity with earthquakes until magnitude 6 caused a large toll in terms of damage and casualties.

We estimated the completeness magnitude for the sequence as 4.3 by using the maximum curvature method. This result is more conservative than the completeness magnitude of 4.0 for the whole Italian catalog from 1960 (Lolli and Gasperini, 2003). Due to the high completeness magnitude, we used the NESTORE_M2 version.

In Figure 7a and b we show the evolution of the sequence in time and the probability (normalized to 1) to have an A cluster estimated by NESTORE_M2 during the first three days of the sequence: all the forecasted probability to have an A cluster are over the 95%. If NESTORE_M2 was available in 1976, the high following seismicity could be forecasted. The alarm could have been available several months before the September destructive swarm, allowing a timely alert to civil protection.

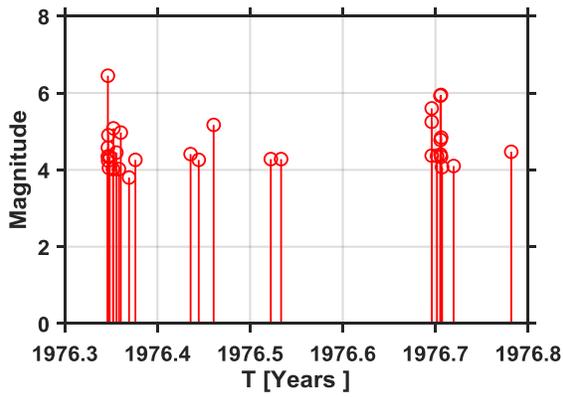

(a)

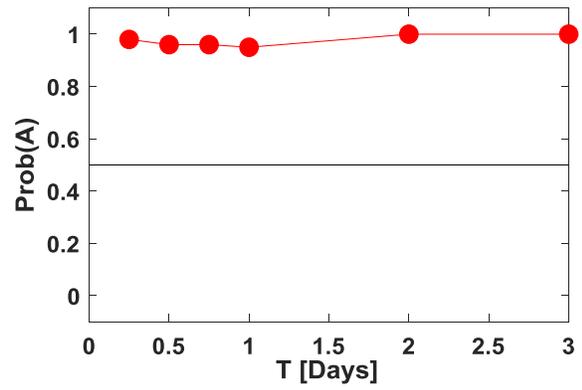

(b)

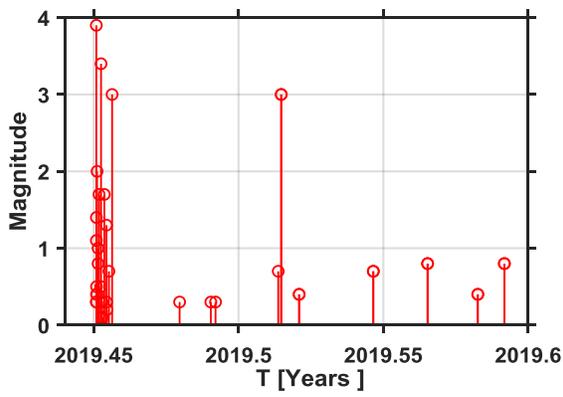

(c)

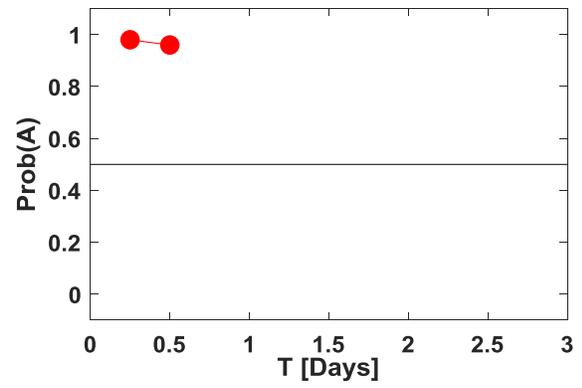

(d)

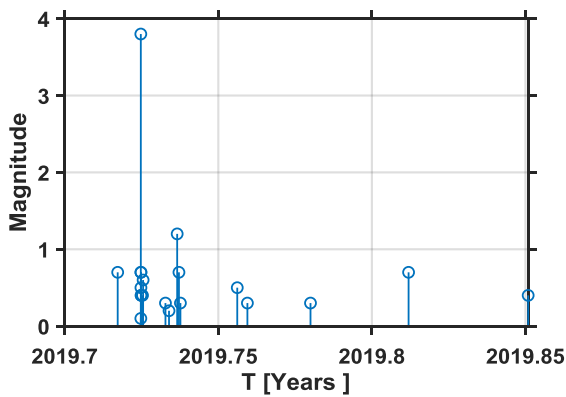

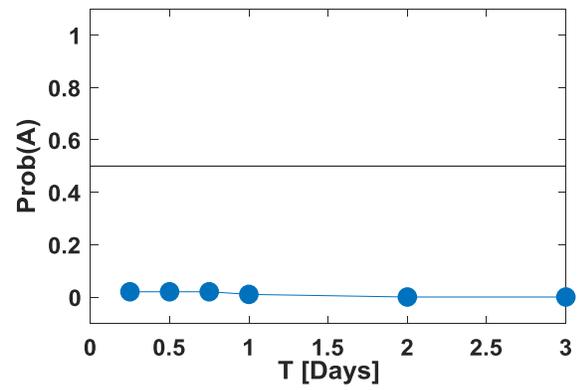

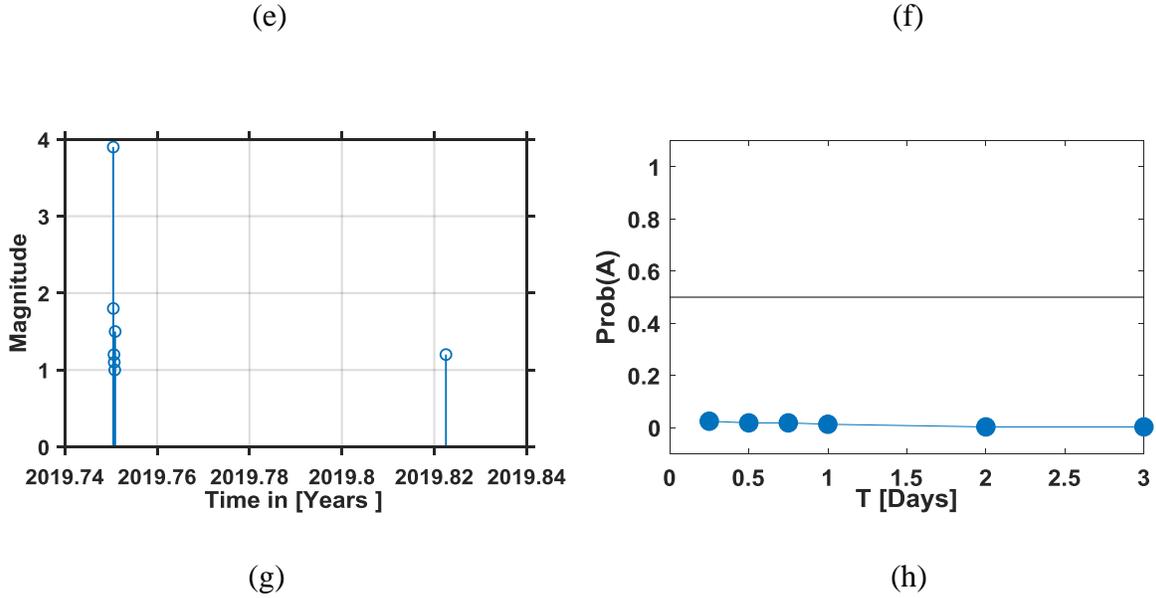

Fig. 7: (a) (c) (e) (g) analyzed clusters. Respectively: Friuli 1976, Tolmezzo June 2019, Tolmezzo September 2019, Ilirska Bistrica (Slovenia) October 2019 (b) (d) (f) (h) Classifications of the clusters by using NESTORE_M2. red: A clusters, blue B clusters.

Figure 7c and d show the time evolution of a small cluster in June 2019 close to the town of Tolmezzo and the probability to have an A cluster forecasted by NESTORE. Even if we used OGS data, NESTORE was trained by data until the end of 2018, so this cluster is not included in the training set. The mainshock of the sequence was on June 14, 2019 M=3.9 and the strongest aftershock of magnitude M=3.4 occurred 14 hours later. Due to the small number of recorded aftershocks, the completeness magnitude could not be evaluated, and it was set to $M_c$=1.5. Also in this case, it was necessary to apply the NESTORE_M2 version of the algorithm. The forecasting is performed until the strongest aftershock arrival i.e. for $T<Dt$. NESTORE_M2 correctly classifies the cluster here too, by supplying a Prob(A)>0.95 for both of the two analyzed time windows.

In Fig. 7e and f another small cluster near Tolmezzo is analyzed, on September 2019. This cluster is type B (mainshock magnitude 3.8) with the strongest aftershock of magnitude 1.2, 4 days after the mainshock (see Fig. 7e). NESTORE_M2 correctly classifies it as a B cluster, by supplying a Prob(A)<0.03.

The last test in Fig 7 g and h and is done by applying NESTORE_M2 to a small cluster of type B in October 2019 near the town of Ilirska Bistrica (Slovenia), in the easternmost part of the studied area. The mainshock of magnitude 3.9 was followed by few aftershocks in the following 4 hours in the magnitude range [1.0, 1.8]. After 26 days with no earthquakes in the area, another small earthquake of magnitude 1.2 was recorded, perhaps not related with the cluster. Here too NESTORE_M2 correctly classifies it as a B cluster, by supplying a Prob(A)<0.03.

Even though these are only four examples of the potentiality of the method, the previous results are encouraging for what concerns the capability of the method to forecast strong aftershocks.

## 5. Summary and discussion

In this paper, we used the OGS seismic catalogue in order to systematically analyze the earthquake clusters with mainshock magnitude $M_m \geq 3.7$ in northeastern Italy and western Slovenia. In particular, we characterized their strongest aftershock magnitude showing the distribution of the difference in magnitude $Dm$ between the mainshock and the strongest aftershock. The distribution is approximately symmetric around the interval [1, 2], with mean 1.3-1.4 in good agreement with Båth's law. The value of $Dm$ increases with $M_m$ until $M_m=5$, while, for larger magnitudes, it decreases, in accordance with the hypothesis of Gentili and Di Giovambattista (2017) that, as larger earthquakes activate more complex tectonic structures, the probability to have a subsequent strong event is higher. However, the dependence on $M_m$ is still debated and other studies on larger magnitude mainshocks found out no dependence (Tahir et al. 2012). Our analysis is performed on a small number of mainshocks with $M_m>5$ and it should be verified on historical database to increase statistical information.

In addition, we analyze the time difference $Dt$ between the mainshock and the strongest aftershock, in order to obtain a regression on the percentage of clusters that already had the strongest aftershock as a function of time elapsed from the mainshock. This information is important for the civil protection purposes, because it suggests how long to wait before to rebuild in security after a strong earthquake. The percentage increases quickly in the first 7 days, while the trend is smoother for longer time periods. We hypothesize a dependence of Dt on the mainshock magnitude, but this result cannot be confirmed by the data, due to the poor statistics.

The main result of this paper is the development of an adaptive pattern recognition method, called NESTORE, for strong forthcoming aftershocks forecasting, based on the approach developed by Gentili and Di Giovambattista (2017). NESTORE is trained by a set of seismological features, which can change from one catalogue to the other, in function of their availability and their reliability, and classifies the cluster in two classes "type A" (if $Dm \leq 1$) or type B (otherwise). In particular, we developed two versions of NESTORE, named NESTORE_M3 and NESTORE_M2; NESTORE_M3 needs a completeness magnitude of the cluster $M_c \leq M_m-3$, like the original method developed by Gentili and Di Giovambattista (2017). NESTORE_M2 uses less features than the previous one and is able to classify clusters with $M_c \leq M_m-2$. In both cases, we developed independent classifiers for each feature separately and we imposed the smallest decision trees (i.e. a threshold on each feature), in order to avoid overfitting due to the small training set available. Only after the training, the resulting classifications are merged using equation (18). If a larger database were available, one only classifier having a pattern of n different features would be used, but in this case, the data are too few to define the n-1 dimension hyperplane separating the two classes, and the problem would be under-constrained.

Thanks to NESTORE_M2 and to the lower completeness magnitude of the catalogue in the OGS region of authority compared to ISIDE data used in Gentili and Di Giovambattista 2017 - from twice to 5 times more earthquakes, depending on the year (Peruzza et al., 2015) - we were able to lower the threshold for application of the pattern recognition method than in Gentili and Di Giovambattista (2017) from $M_m=4.5$ to $M_m=3.7$.

We evaluated the feature performances independently, using one-node decision trees (a threshold method) for classification, and LOO method for performances evaluation. In particular, we analyzed Precision, Recall, Accuracy and Informedness of the features, eliminating from NESTORE the features with negative or zero Informedness and the ones with Accuracy lower than the one of a constant response classifier. The remaining ten (seven for NESTORE_M2) features performances were visualized by ROC diagrams and turned out to be well over the diagonal corresponding to random response. We proposed an innovative method for combining the different classifiers based on Bayes approach, which takes into account the different reliability of each feature if its value is over or under the selected threshold.

In order to identify the outliers, we re-classified the training set. An analysis evidenced six outliers, showing features values different from the ones of the corresponding class. Gentili and Di Giovambattista (2017) found an outlier for the method during the analysis of Central Italy seismicity: it was Gualdo Tadino cluster, which occurred on 3 April 1998, in a positive stress area at the edge of the seismogenic volume activated by the 26 September 1997 Mw 6.0 mainshock of the Umbria-Marche sequence. A similar mechanism may explain the anomalous behavior of the seismicity of Trasaghis 98 and Kobarid 99 clusters (a type B cluster classified as A and an A cluster classified as B, respectively). The Kobarid 98 (mainshock 5.6) cluster, close in time and space, may have influenced both of them. The 1976 Friuli sequence and a small cluster in Trasaghis few months before, similarly, may have influenced Trasaghis 77 cluster. However, the classification of this cluster becomes correct after two days. An alternative explanation may be that, since this was the first cluster recorded by OGS in 1977 few months after the installation of the seismic network, when few seismic stations were available (see Gentili et al., 2011), the incorrect classification in the first day may be related to inaccuracy in earthquake recording. Other two wrongly classified A clusters are Valdobbiadene 15 and Forni di Sopra in 18. For Valdobbiadene 15 we found a cluster in the same area 4 years before (Valdobbiadene 11) but the magnitude of the maishock is small ($M_L$=3.7) and, in our opinion, a long term-interaction of the two earthquake clusters is unlikely. A different analysis is necessary for Gemona 78, a type B cluster wrongly classified as A. In that case, not all the features match in the classification as A, because Z, SLcum and QLcum decision trees classify it as B. If we use NESTORE_M3, the classification becomes correct, due to the contribution of N, $V_m$ and $V_n$ features. Unfortunately, due to too high completeness magnitude, only Gemona 78 and Trasaghis 77 outliers could be classified using NESTORE_M3.

As a final test, we applied NESTORE_M2 to four clusters not contained in the training set. NESTORE_M2 supplied good results for 1976 Friuli A type earthquake cluster, which struck the analyzed region before the starting date of the OGS catalogue causing many casualties also because of the unexpected strong events four months after the mainshock. Other three successful classifications were two small clusters (an A and a B cluster) near Tolmezzo and one near Ilirska

Bistrica (Slovenia) in 2019, not included in the training set. There is the need to test NESTORE in a fully prospective sense. Tests started with Tolmezzo and Ilirska Bistrica clusters and will need several years to reach a meaningful statistical power. The decrease of completeness magnitude in the last years due to a larger coverage of the region with seismic stations and to collaboration with ARSO, however, will shorten the times to reach such test set. NESTORE will be continuously tested on future seismicity in the next years to verify its performances, with the goal of adopting the method for decision-making during seismic crises.

**Acknowledgments.** The OGS seismic network is managed by the Sezione Centro di Ricerche Sismologiche of OGS and financially supported by the Civil Protection of the Regione Autonoma Friuli Venezia Giulia, Italy and of the Regione Veneto, Italy and by the Geological Survey of Provincia Autonoma di Trento, Italy. We thank the technical staff of the Department Centro di Ricerche Sismologiche (OGS) for data acquisition and processing.

## *Appendix*

In pattern recognition applications, binary classifier performances are summarized in a matrix called *Confusion matrix* (or *contingency table*) – see Fig. 1A. Defining one class as positive (we choose A class) and the other as negative (we choose B class) and testing the classifier on a set of examples of known classification, it is possible to evaluate the number of correct classifications and mistakes in dependence of the correct class.

|  | True class | |
|---|---|---|
|  | p | n |
| **Y** (Hypothesized) | True Positive (TP) | False Positive (FP) |
| **N** | False Negative (FN) | True Negative (TN) |
| Column totals | P | N |

Fig. 1A: Confusion Matrix

In particular, four outcomes of classification are stored in the matrix:

- TP= True Positives : positive examples correctly classified as positive
- TN= True Negatives: negative examples correctly classified as negative
- FN=False Negatives: positive examples wrongly classified as negatives
- FP= False Positives: negative examples wrongly classified as positive

The numbers along the major diagonal represent the correct decisions made, while the numbers off this diagonal represent the errors—the confusion—between the various classes.

*Recall*, *Precision*, *Accuracy* and *Informedness* use the *Confusion Matrix* elements to infer some information on the classifier.

The *Recall*, also called *True Positive Rate*, is a measure of the classifier ability to identify positives as positives. It is defined as:

$$\text{Recall} = \frac{\text{Positives correctly classified}}{\text{Positives}} = \frac{TP}{P} = \frac{TP}{TP+FN} \quad (1)$$

The *Precision* is a measure of the trustworthiness of the positive prediction and is defined as:

$$\text{Precision} = \frac{\text{Positives correctly classified}}{\text{Classified as positive}} = \frac{TP}{Y} = \frac{TP}{TP+FP} \quad (2)$$

The *Accuracy* is a measure of the percentage of correct classifications considering both classes and is defined as:

$$\text{Accuracy} = \frac{\text{Correctly classified}}{\text{All}} = \frac{TP+TN}{P+N} = \frac{TP+TN}{TP+FP+FN+TN} \quad (3)$$

The *Informedness* is a measure on how informed is the classifier about the class and is defined as the difference between the *Recall* and the *False Positive Rate* where:

$$\text{False Positive Rate} = \frac{\text{Negatives incorrectly classified}}{\text{Total negatives}} = \frac{FP}{N} = \frac{FP}{FP+TN} \quad (4)$$

*Precision*, *Recall*, and *Accuracy* range between 0 and 1 (0=the worst classifier 1=the best classifier) Informedness ranges between -1 and 1 (-1=the worst classifier 1=the best classifier). The binary classifier performances are usually visualized by a Receiver Operating Characteristics (ROC) graph (Egan, 1975; Swets et al., 2000) which shows True Positive Rate against the False Positive Rate. A ROC graph depicts relative trade-offs between benefits (true positives) and costs (false positives) (see Fig. 2A).

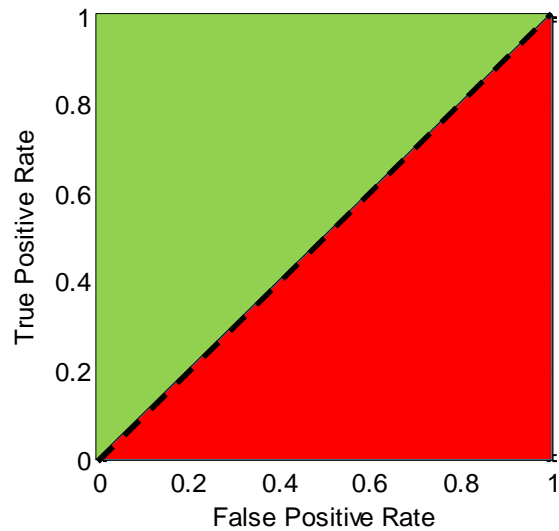

Figure 2A: ROC graph.

A discrete classifier produces a point of coordinates (False Positive Rate, True Positive Rate) in the ROC graph. Diagonal (dashed line) represents random guessing. The points over the dashed line (green triangle) correspond to a classifier better that random guessing, the points under to a classifier (red triangle) worse than random guessing. One point in ROC space is better than another if it is closer to the point (0,1) (perfect classifier). *Informedness*, being the difference between True Positive Rate and False Positive Rate, is a measure of the distance from the perfect classifier. Further details can be found in Gentili and Di Giovambattista (2017).